\documentclass[prd,superscriptaddress,nofootinbib,amsmath,amssymb,aps,11pt]{revtex4}

\usepackage{bm}
\usepackage{amsfonts}
\usepackage{latexsym}
\usepackage[latin1]{inputenc}
\usepackage{graphicx}
\usepackage{amsmath}
\usepackage{palatino}
\usepackage{mathpazo}
\usepackage{booktabs}
\usepackage{dcolumn}

\def\jnl@style{\it}
\def\aaref@jnl#1{{\jnl@style#1}}

\def\aaref@jnl#1{{\jnl@style#1}}

\def\aj{\aaref@jnl{AJ}}                   % Astronomical Journal
\def\apj{\aaref@jnl{ApJ}}                 % Astrophysical Journal
\def\apjl{\aaref@jnl{ApJ}}                % Astrophysical Journal, Letters
\def\apjs{\aaref@jnl{ApJS}}               % Astrophysical Journal, Supplement
\def\apss{\aaref@jnl{Ap\&SS}}             % Astrophysics and Space Science
\def\aap{\aaref@jnl{A\&A}}                % Astronomy and Astrophysics
\def\aapr{\aaref@jnl{A\&A~Rev.}}          % Astronomy and Astrophysics Reviews
\def\aaps{\aaref@jnl{A\&AS}}              % Astronomy and Astrophysics, Supplement
\def\mnras{\aaref@jnl{Mon.~Not.~Roy.~Astron.~Soc.}}             % Monthly Notices of the RAS
\def\prd{\aaref@jnl{Phys.~Rev.~D}}        % Physical Review D
\def\prc{\aaref@jnl{Phys.~Rev.~C}}  % Physical Review C
\def\prl{\aaref@jnl{Phys.~Rev.~Lett.}}    % Physical Review Letters
\def\qjras{\aaref@jnl{QJRAS}}             % Quarterly Journal of the RAS
\def\skytel{\aaref@jnl{S\&T}}             % Sky and Telescope
\def\ssr{\aaref@jnl{Space~Sci.~Rev.}}     % Space Science Reviews
\def\zap{\aaref@jnl{ZAp}}                 % Zeitschrift fuer Astrophysik
\def\nat{\aaref@jnl{Nature}}              % Nature
\def\aplett{\aaref@jnl{Astrophys.~Lett.}} % Astrophysics Letters
\def\apspr{\aaref@jnl{Astrophys.~Space~Phys.~Res.}} % Astrophysics Space Physics Research
\def\physrep{\aaref@jnl{Phys.~Rep.}}      % Physics Reports
\def\physscr{\aaref@jnl{Phys.~Scr}}       % Physica Scripta
\def\commat{\aaref@jnl{Comm.~Math.~Phys.}}              % Communications in Mathematical Physics
\def\science{\aaref@jnl{Science}}               % Science
\def\cqg{\aaref@jnl{Classical Quant.~Grav.}}            % Classical and Quantum Gravity
\def\jpcs{\aaref@jnl{JPCS}}                                     % Journal of Physics Conference Series
\def\ijmpd{\aaref@jnl{Int.~J.~Mod.~Phys.~D}}                    % International Journal of Modern Physics D
\def\grg{\aaref@jnl{Gen.~Relat.~Gravit.}}               % General Relativity and Gravitation
\def\rpp{\aaref@jnl{Rep.~Prog.~Phys.}}          % Reports on Progress in Physics
\def\npa{\aaref@jnl{Nucl.~Phys.~A}}        % Nuclear Physics A
\def\lrr{\aaref@jnl{Living Rev.~Rel.}}                   % Living reviews in relativity
\def\jcap{\aaref@jnl{J.~Cosmology Astropart.~Phys.}}    % Journal of cosmology and astroparticle physics
\def\rmp{\aaref@jnl{Rev.~Mod.~Phys.}}   %Reviews of modern physics

\allowdisplaybreaks[1]
\begin{document}

\title{Gauss-Bonnet black holes with a massive scalar field}

\author{Daniela D. Doneva}
\email{daniela.doneva@uni-tuebingen.de}
\affiliation{Theoretical Astrophysics, Eberhard Karls University of T\"ubingen, T\"ubingen 72076, Germany}
\affiliation{INRNE - Bulgarian Academy of Sciences, 1784  Sofia, Bulgaria}

\author{Kalin V. Staykov}
\email{kstaykov@phys.uni-sofia.bg}
\affiliation{Department of Theoretical Physics, Faculty of Physics, Sofia University, Sofia 1164, Bulgaria}

\author{Stoytcho S. Yazadjiev}
\email{yazad@phys.uni-sofia.bg}
\affiliation{Theoretical Astrophysics, Eberhard Karls University of T\"ubingen, T\"ubingen 72076, Germany}
\affiliation{Department of Theoretical Physics, Faculty of Physics, Sofia University, Sofia 1164, Bulgaria}
\affiliation{ Institute of Mathematics and Informatics, Bulgarian Academy of Sciences,
	Acad. G. Bonchev Street 8, Sofia 1113, Bulgaria}

%%%%%%%%%%%%%%%%%%%%%%%%%%%%%%%%%%%%  DATE  %%%%%%%%%%%%%%%%%%%%%%%%%%%%%%%%%%%%
%\date{\today}

\begin{abstract}
In the present paper we consider the extended scalar-tensor-Gauss-Bonnet gravity with a massive scalar field. We prove numerically the existence of Gauss-Bonnet black holes for three different forms of the coupling function including the case of spontaneous scalarization. We have performed a systematic study of the black hole characteristics such as the area of the horizon, the entropy and the temperature for these coupling functions and compared them to the Schwarzschild solutions. The introduction of scalar field mass leads to a suppression of the scalar field and the increase of this mass brings the black holes closer to the Schwarzschild case.  For linear and exponential coupling, a nonzero scalar field mass expands the domain of existence of black holes solutions. Larger deviations from the Schwarzschild solution are observed only for small masses and these differences decrease with the increase of the scalar field mass. In the case of a coupling function which leads to scalarization the scalar field mass has a significant influence on the bifurcation points where the scalarized black holes branch out of the Schwarzschild solution. The largest deviation from the case with a massless scalar field  are observed for black hole masses close to the bifurcation point. 
\end{abstract}

%\pacs{04.40.Dg, 04.50.Kd, 04.80.Cc}

\maketitle

\section{Introduction}

Ones of the natural modifications of general relativity (GR) are the extended scalar-tensor theories (ESTT) where the usual Einstein-Hilbert action is supplemented with all possible algebraic curvature invariants of second order with a dynamical scalar field nonminimally coupled to these invariants \cite{Berti_2015}--\cite{Pani_2011a}. A particular sector of ESTT is the extended scalar-tensor-Gauss-Bonnet (ESTGB) gravity for which the scalar field is coupled exactly to the topological Gauss-Bonnet invariant. The field equations of the  ESTGB gravity are of second order as in general relativity
contrary to the general ESTT where they are of higher order.  One of the most studied models in the last decade within ESTGB gravity  is the so-called Einstein-dilaton-Gauss-Bonnet (EdGB) gravity which is characterized by the coupling function  $\alpha e^{\gamma \varphi}$ for the dilaton field, with  $\alpha$ and $\gamma$ being constants. The nonrotating  EdGB black holes were studied perturbatively or numerically in \cite{Mignemi_1993}--\cite{Pani_2009}. It was shown that the EdGB black holes exist when the black hole mass is greater than certain lower bound proportional to the parameter $\alpha$. The slowly rotating black holes in EdGB gravity were studied  in \cite{Pani_2009}, \cite{Ayzenberg_2014} and \cite{Maselli_2015}. The rapidly rotating EdGB black holes were constructed
numerically in \cite{Kleihaus_2011}--\cite{Kleihaus:2015aje}. The rotating EdGB black holes can exist only when the mass and the angular momentum fall in certain domain depending on the coupling constant. Another very interesting fact  about the EdGB black holes is that they can exceed the Kerr bound for the angular momentum.  The ESTGB black holes were further studied in  \cite{Antoniou_2018}--\cite{Chen:2016qks}. The stability and the quasinormal modes of EdGB black holes were examined in \cite{Blazquez-Salcedo:2016enn,Blazquez-Salcedo:2017txk}. The dynamical evolution in Gauss-Bonnet gravity and different aspects of collapse were examined  in  \cite{Benkel:2016rlz}--\cite{Chakrabarti:2017apq}.

Recently the interest in ESTGB gravity was provoked  by the discovery of the  spontaneous scalarization of  the Schwarzschild  black holes within a certain class of ESTGB gravity \cite{Doneva_2018,Silva_2018}. It was shown that in a certain class of ESTGB theories there exist new black hole solutions which are formed by spontaneous scalarization of the Schwarzschild black holes in the extreme curvature regime. In this regime, below certain mass, the Schwarzschild solution becomes unstable and a new branch of solutions with nontrivial scalar field bifurcate from the Schwarzschild one. As a matter of fact, more than one branches with nontrivial scalar field can bifurcate at different masses but only the first one can be stable. In contrast with the standard spontaneous scalarization of neutron stars \cite{Damour_1993} and black holes \cite{Stefanov_2008, Doneva_2010} in the standard scalar-tensor theories, which is induced by the presence of matter, in the  case under consideration  the scalarization is induced by the curvature of the spacetime. 
The spontaneous scalarization and the scalarized black holes in ESTGB theories were studied in different aspects in many papers \cite{Myung_2018}-\cite{Brihay_2019b}.

The ESTGB black holes have been studied only in the case when the scalar is massless. As a first step this is quite a natural assumption. However, more realistic treatment of the problem requires a massive scalar field and even  a scalar field with  self-interaction as done in \cite{Staykov:2018hhc} for the case of neutron stars. Indeed, certain sectors of extended scalar-tensor theories arise naturally in string
theory and at low energies supersymmetry is broken which leads to a massive scalar field. The inclusion of scalar field mass can change the picture considerably. It suppresses the scalar field at length scale of the order of the Compton wavelength which helps us reconcile the theory
with the observations for a much broader range of the coupling parameters and functions.   

The purpose  of the present paper is to study the black holes in ESTGB gravity with a massive
scalar field. More precisely we construct numeral black holes solutions in ESTGB gravity with linear and exponential coupling for the scalar field and also study their basic properties. We also study  the spontaneous scalarization of Schwarzschild black hole in ESTGB gravity  and show that there exist scalarized Gauss-Bonnet black holes with a massive scalar field. Some basic characteristics of the 
Gauss-Bonnet black holes with a massive scalar field are studied too.

\section{Basic equations and setting the problem}

The general vacuum action of ESTGB theories   is described  by  
\begin{eqnarray}\label{GBA}
S=&&\frac{1}{16\pi}\int d^4x \sqrt{-g} 
\Big[R - 2\nabla_\mu \varphi \nabla^\mu \varphi - V(\varphi) 
 + \lambda^2 f(\varphi){\cal R}^2_{GB} \Big] ,\label{eq:quadratic}
\end{eqnarray}
where $R$ is the Ricci scalar curvature with respect to the spacetime metric $g_{\mu\nu}$, $\varphi$ is the scalar field with a potential $V(\varphi)$ and a coupling function  $f(\varphi)$ depending only on $\varphi$, $\lambda$ is the Gauss-Bonnet coupling constant having  dimension of $length$ and ${\cal R}^2_{GB}$ is the Gauss-Bonnet invariant\footnote{The Gauss-Bonnet invariant is defined by ${\cal R}^2_{GB}=R^2 - 4 R_{\mu\nu} R^{\mu\nu} + R_{\mu\nu\alpha\beta}R^{\mu\nu\alpha\beta}$ where $R$ is the Ricci scalar, $R_{\mu\nu}$ is the Ricci tensor and $R_{\mu\nu\alpha\beta}$ is the Riemann tensor}.  The field equations derived by the action (\ref{GBA}) are the following  
\begin{eqnarray}\label{FE}
&&R_{\mu\nu}- \frac{1}{2}R g_{\mu\nu} + \Gamma_{\mu\nu}= 2\nabla_\mu\varphi\nabla_\nu\varphi -  g_{\mu\nu} \nabla_\alpha\varphi \nabla^\alpha\varphi - \frac{1}{2} g_{\mu\nu}V(\varphi), \notag\\
&&\nabla_\alpha\nabla^\alpha\varphi= \frac{1}{4} \frac{dV(\varphi)}{d\varphi} -  \frac{\lambda^2}{4} \frac{df(\varphi)}{d\varphi} {\cal R}^2_{GB}, \label{FE}
\end{eqnarray}
where  $\nabla_{\mu}$ is the covariant derivative with respect to the spacetime metric $g_{\mu\nu}$ and  $\Gamma_{\mu\nu}$ is defined by 

\begin{eqnarray}
&&\Gamma_{\mu\nu}= - R(\nabla_\mu\Psi_{\nu} + \nabla_\nu\Psi_{\mu} ) - 4\nabla^\alpha\Psi_{\alpha}\left(R_{\mu\nu} - \frac{1}{2}R g_{\mu\nu}\right) + 
4R_{\mu\alpha}\nabla^\alpha\Psi_{\nu} + 4R_{\nu\alpha}\nabla^\alpha\Psi_{\mu} \nonumber \\ 
&& - 4 g_{\mu\nu} R^{\alpha\beta}\nabla_\alpha\Psi_{\beta} 
 + \,  4 R^{\beta}_{\;\mu\alpha\nu}\nabla^\alpha\Psi_{\beta} 
\end{eqnarray}  
with 
\begin{eqnarray}
\Psi_{\mu}= \lambda^2 \frac{df(\varphi)}{d\varphi}\nabla_\mu\varphi .
\end{eqnarray}
In the preset paper we shall focus on the simplest massive potential, namely   
\begin{equation}\label{eq:Potential}
V(\varphi)=2 m_\varphi^2\varphi^2 ,
\end{equation}
where $m_\varphi$ is the mass of the scalar field.

We consider  static and spherically symmetric spacetimes as well as static and spherically symmetric 
scalar field configurations. The spacetime metric  can then be written in the standard form
\begin{eqnarray}
ds^2= - e^{2\Phi(r)}dt^2 + e^{2\Lambda(r)} dr^2 + r^2 (d\theta^2 + \sin^2\theta d\phi^2 ). 
\end{eqnarray}   
The dimensionally reduced field equations (\ref{FE}) are the following 
\begin{eqnarray}
&&\frac{2}{r}\left[1 +  \frac{2}{r} (1-3e^{-2\Lambda})  \Psi_{r}  \right]  \frac{d\Lambda}{dr} + \frac{(e^{2\Lambda}-1)}{r^2} 
- \frac{4}{r^2}(1-e^{-2\Lambda}) \frac{d\Psi_{r}}{dr} - \left( \frac{d\varphi}{dr}\right)^2 - \frac{1}{2}V(\varphi)e^{2\Lambda}=0, \label{DRFE1}\\ && \nonumber \\
&&\frac{2}{r}\left[1 +  \frac{2}{r} (1-3e^{-2\Lambda})  \Psi_{r}  \right]  \frac{d\Phi}{dr} - \frac{(e^{2\Lambda}-1)}{r^2} - \left( \frac{d\varphi}{dr}\right)^2 + \frac{1}{2}V(\varphi)e^{2\Lambda}=0,\label{DRFE2}\\ && \nonumber \\
&& \frac{d^2\Phi}{dr^2} + \left(\frac{d\Phi}{dr} + \frac{1}{r}\right)\left(\frac{d\Phi}{dr} - \frac{d\Lambda}{dr}\right)  + \frac{4e^{-2\Lambda}}{r}\left[3\frac{d\Phi}{dr}\frac{d\Lambda}{dr} - \frac{d^2\Phi}{dr^2} - \left(\frac{d\Phi}{dr}\right)^2 \right]\Psi_{r} 
\nonumber \\ 
&& \hspace{0.5cm} - \frac{4e^{-2\Lambda}}{r}\frac{d\Phi}{dr} \frac{d\Psi_r}{dr} + \left(\frac{d\varphi}{dr}\right)^2 + \frac{1}{2}V(\varphi)e^{2\Lambda}=0, \label{DRFE3}\\ && \nonumber \\
&& \frac{d^2\varphi}{dr^2}  + \left(\frac{d\Phi}{dr} \nonumber - \frac{d\Lambda}{dr} + \frac{2}{r}\right)\frac{d\varphi}{dr} -\frac{1}{4}\frac{dV(\varphi)}{d\varphi}e^{2\Lambda} \nonumber \\ 
&& \hspace{0.5cm} - \frac{2\lambda^2}{r^2} \frac{df(\varphi)}{d\phi}\Big\{(1-e^{-2\Lambda})\left[\frac{d^2\Phi}{dr^2} + \frac{d\Phi}{dr} \left(\frac{d\Phi}{dr} - \frac{d\Lambda}{dr}\right)\right]    + 2e^{-2\Lambda}\frac{d\Phi}{dr} \frac{d\Lambda}{dr}\Big\}  =0, \label{DRFE4}
\end{eqnarray}
with 
\begin{eqnarray}
\Psi_{r}=\lambda^2 \frac{df(\varphi)}{d\varphi} \frac{d\varphi}{dr}.
\end{eqnarray}

 The boundary and the regularity conditions for the above system of differential equations are as follows. As usual the asymptotic flatness imposes the following asymptotic conditions
\begin{eqnarray}
\Phi|_{r\rightarrow\infty} \rightarrow 0, \;\;  \Lambda|_{r\rightarrow\infty} \rightarrow 0,\;\; \varphi|_{r\rightarrow\infty} \rightarrow 0\;\;.   \label{eq:BH_inf}
\end{eqnarray} 
The very existence of black hole horizon at $r=r_H$ requires 
\begin{eqnarray}
e^{2\Phi}|_{r\rightarrow r_H} \rightarrow 0, \;\; e^{-2\Lambda}|_{r\rightarrow r_H} \rightarrow 0. \label{eq:BC_rh}
\end{eqnarray} 

The regularity of the scalar field and its first  and second derivatives on the black hole horizon
gives one more condition. From this condition  one can derive the following  condition for existence of black hole solutions, namely
\begin{eqnarray}\label{QE}
&&16 (m_{\varphi}\lambda)^2 \left(\frac{df(\varphi)}{d\varphi}\right)^2_{H}\varphi_{H} \left\{\left[(m_{\varphi}r_H)^2\varphi_H^3-6\right]\left(\frac{df(\varphi)}{d\varphi}\right)^2_{H}  
+ 3 \left(\frac{df(\varphi)}{d\varphi}\right)_H \left(\frac{r_H}{\lambda}\right)^2  + \varphi_{H} \left(\frac{r_H}{\lambda}\right)^4 \right\} \nonumber \\ 
&& + \left(\frac{r_H}{\lambda}\right)^6 - 24 \left(\frac{df(\varphi)}{d\varphi}\right)^2_H\left(\frac{r_H}{\lambda}\right)^2  \ge 0 .
\end{eqnarray}

The mass of the black hole $M$ is obtained through the asymptotics of the function $\Lambda$ or $\Phi$, namely 
\begin{equation}
\Lambda\approx \frac{M}{r} + O(1/r^2), \;\; \Phi\approx  -\frac{M}{r} + O(1/r^2).
\end{equation}
Concerning the asymptotic of the scalar field we have 
\begin{equation}
\varphi\sim \frac{e^{-mr}}{r}. 
\end{equation}

\section{Numerical results}

The the system (\ref{DRFE1})--(\ref{DRFE4}) is solved numerically using a shooting method that is discussed in detail in \cite{Doneva_2018}. The difference with the present case is the addition of scalar field mass that makes the system of equations (\ref{DRFE1})--(\ref{DRFE4}) stiff. This complicates the numerical solution of the problem significantly and requires careful adjustments of the parameters of the code \cite{Yazadjiev:2016pcb,Staykov:2018hhc}.  

We will consider three different forms of the coupling function $f(\varphi)$. The first two cases are  a linear and  an exponential coupling. The third coupling function is the function that allows for spontaneous scalarization of the Schwarzschild black hole
in the ESTGB gravity with a massless scalar field \cite{Doneva_2018}. 

\subsection{Black holes with linear coupling }

We start our study with the simplest case of linear coupling function
\begin{equation}\label{eq:linear_coupling}
f(\varphi) = \varphi
\end{equation} 
and potential having the form of eq. \eqref{eq:Potential}. Thus the free parameters of the problem are the mass of the scalar field and the parameter $\lambda$. The quantities presented below, such as the black hole mass, radius of the horizon, etc., are scaled with respect to $\lambda$.

\begin{figure}[]
	\centering
	\includegraphics[width=0.45\textwidth]{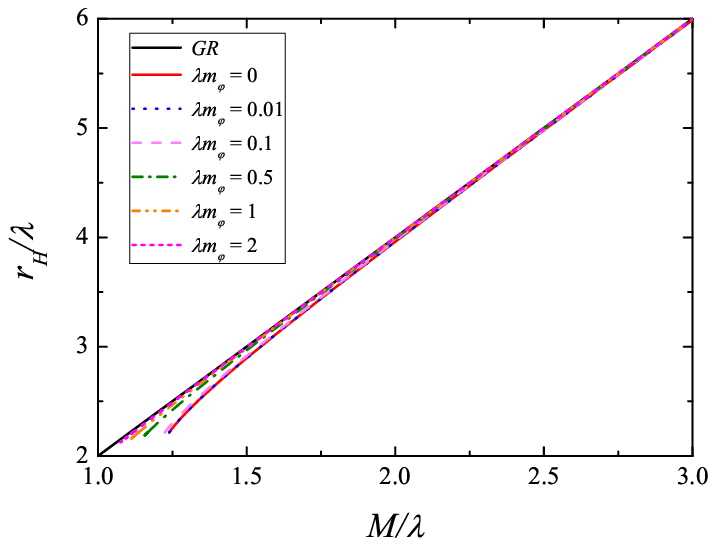}
	\includegraphics[width=0.45\textwidth]{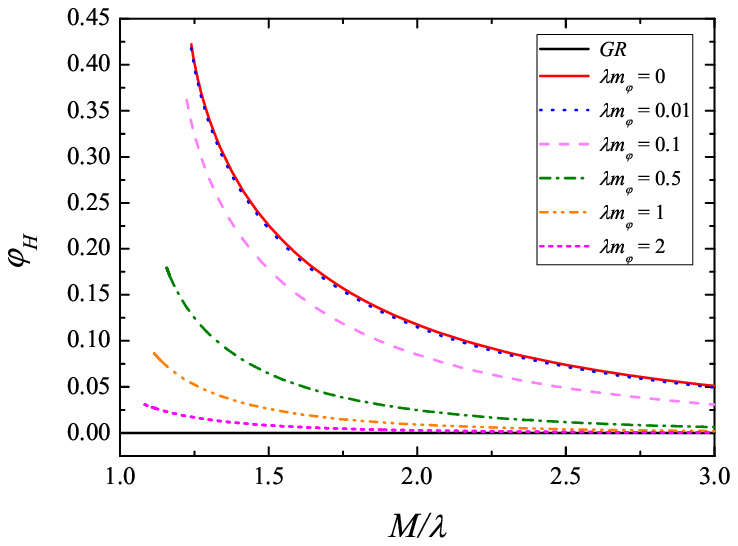}
	\caption{\textit{Left} The radius of the horizon versus the black hole mass for the linear coupling function \eqref{eq:linear_coupling}. Both are rescaled with respect to the coupling constant $\lambda$. \textit{Right} The value of the scalar field on the horizon versus the rescaled black hole mass. The notations on both panels are identical. The results are for different values for the scaled mass of the scalar field $\lambda m_{\varphi}$ in different colours and patterns. }
	\label{Fig:rh_phi}
\end{figure}

In the \textit{left} panel of Fig. \ref{Fig:rh_phi} we plot the black hole radius versus the black hole mass for different values of 
the scalar field mass $m_\varphi$.  All these quantities are rescaled with respect to the coupling constant $\lambda$ as we commented and for simplicity, hereafter we will refer to them without mentioning the scaling explicitly.  One can see that  the radius of the black hole is always smaller, compared to the GR case. The deviation is maximal for small black hole masses and decreases with the increase of the mass of the black hole. When the massless field is exchanged with a massive one, it is clear that with the increase of the mass of the field, the radius converge to the Schwarzschild one and the scalar field decreases as well. This is completely expected since introducing a mass of the scalar field leads to the fact that the scalar field is loosely speaking confined within its Compton wavelength and higher masses lead to smaller Compton wavelengths.  For each  sequence of Gauss-Bonnet black holes  there exists a minimal mass $M_{\rm crit}$ below which no  black hole solutions exist because condition (\ref{QE}) is violated.  Introducing a scalar field mass leads to a decrease of $M_{\rm crit}$, i.e. the domain of existence of black hole solutions expands and larger scalar field mass leads to smaller  $M_{\rm crit}$. 

In the \textit{right} panel of Fig. \ref{Fig:rh_phi} we present the value of the scalar field on the horizon $\varphi_H$ as a function of the black hole mass. It is clear that $\varphi_H$ decreases rapidly with the increase of the black hole mass which is in accordance with the larger deviations from the Schwarzschild solution in the black hole horizon for small masses $M$. In this figure we can observe as well the suppression of the scalar field with the increase of $m_\varphi$, that we commented above.

\begin{figure}[]
	\centering
	\includegraphics[width=0.45\textwidth]{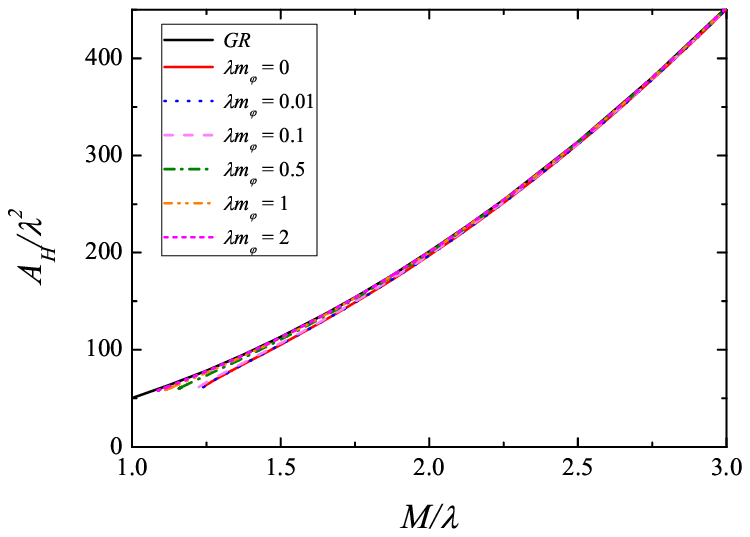}
	\includegraphics[width=0.45\textwidth]{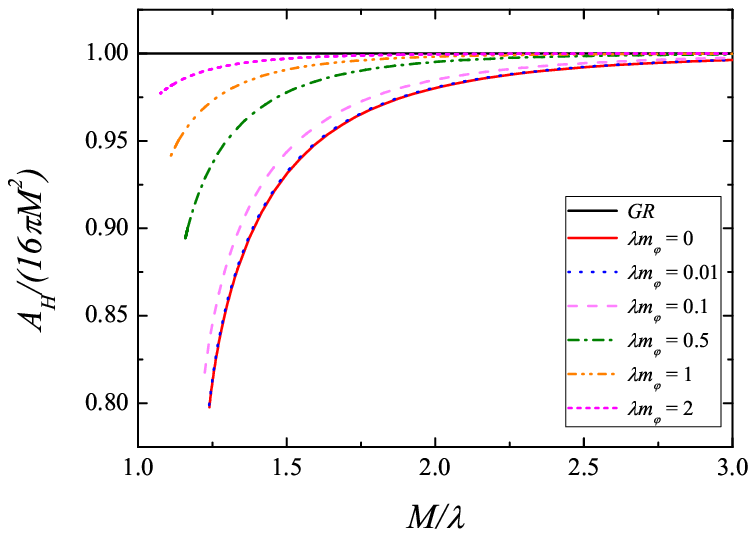}
	\caption{\textit{Left} The area $A_H$ of the black hole horizon versus the mass of the black hole for the linear coupling function \eqref{eq:linear_coupling}. Both are rescaled with respect to the coupling constant $\lambda$. \textit{Right} The normalized, to the Schwarzschild limit, arena of the horizon $A_{H}/(16\pi M^2)$ versus the rescaled mass of the black hole. The notations in both panels are identical. The results are for different values for the scaled mass of the scalar field $\lambda m_{\varphi}$  in different colours and patterns. }
	\label{Fig:Ah_phi}
\end{figure}

In the \textit{left} panel of Fig. \ref{Fig:Ah_phi}  we plot the area of the horizon, $A_H = 4\pi r_H^2$, rescaled by $\lambda$, versus the mass of the black hole. In this case as well the deviations from GR are larger only for the smallest black hole masses for which the solutions in ESTGB gravity exist, and they converge to GR when the scalar field mass $m_\varphi$ increases. In the \textit{right} panel we plot the the area of the horizon, normalized to the Schwarzschild area $A_{H}/(16\pi M^2)$, which gives us a better representation of the deviations from GR. The normalized area of the horizon is smaller for the massless case, and it increases with the increase of the scalar field mass, converging to the GR case. 

\begin{figure}[]
	\centering
	\includegraphics[width=0.45\textwidth]{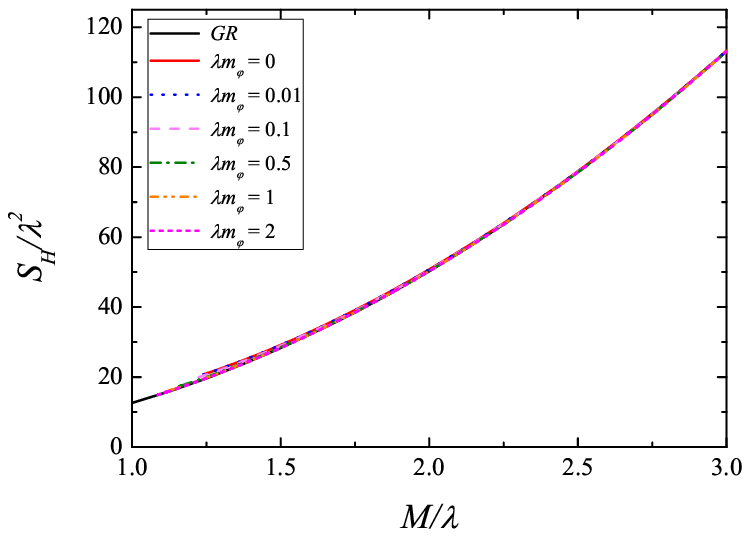}
	\includegraphics[width=0.45\textwidth]{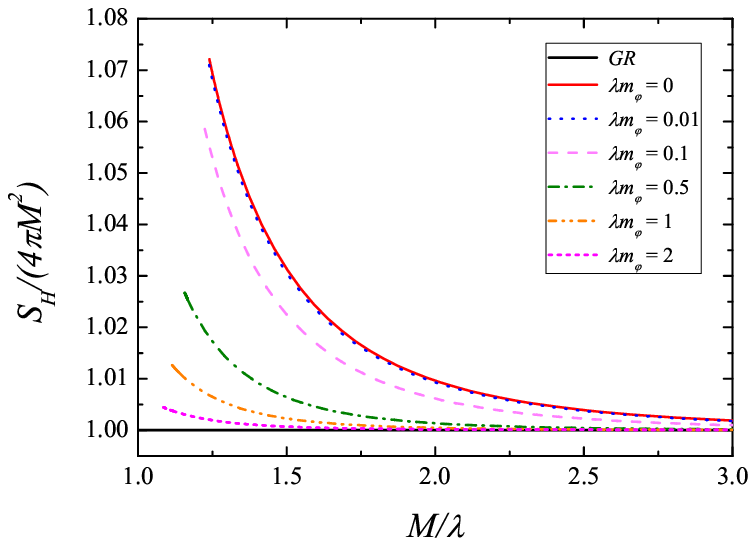}
	\caption{\textit{Left} The entropy of the black hole versus its mass for the linear coupling function \eqref{eq:linear_coupling}. Both are rescaled to the coupling constant. \textit{Right} The normalized to the Schwarzschild limit entropy on the horizon, $S_H/(4\pi M^2)$, versus the rescaled mass of the black hole. The notations on both panels are the same as in the figures above. }
	\label{Fig:Sh_phi}
\end{figure}

In the \textit{left} panel of Fig. \ref{Fig:Sh_phi} we plot the rescaled entropy of the black hole as a function of its mass. We adopt the entropy 
formula proposed by  Wald  \cite{Wald_1993,Iyer_1994}, namely

\begin{equation}\label{eq:S_linear_coupling}
S_H = \frac{1}{4}A_H + 4 \pi\lambda^2 f(\varphi_{H}).
\end{equation}

In this case as well, minor deviations are observed increasing only for small masses. For better presentation of the deviations from GR in the \textit{right} panel we plot the normalized to the Schwarzschild limit entropy, $S_{H}/(4\pi M^2)$ versus the black hole mass. In this normalization it is clear that the entropy of the Gauss-Bonnet black holes is always larger than the entropy of the Schwarzschild black hole, and it decreases with the increase of the mass of the scalar field. 

\begin{figure}[]
	\centering
	\includegraphics[width=0.45\textwidth]{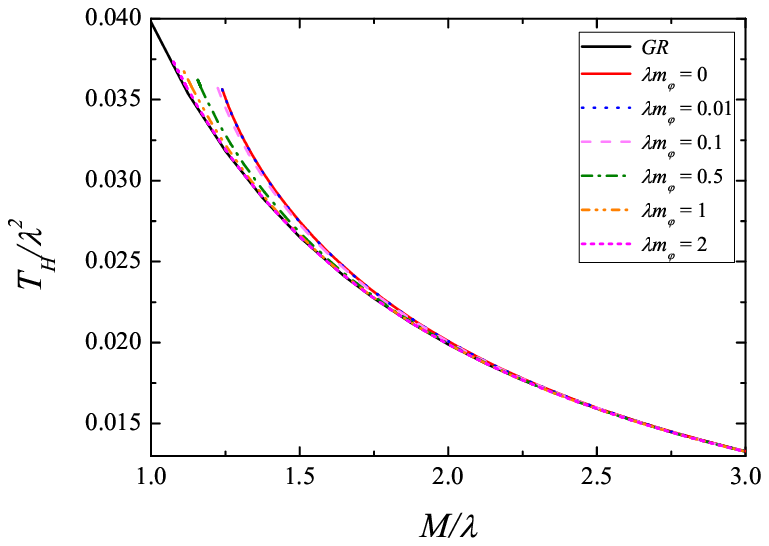}
	\includegraphics[width=0.45\textwidth]{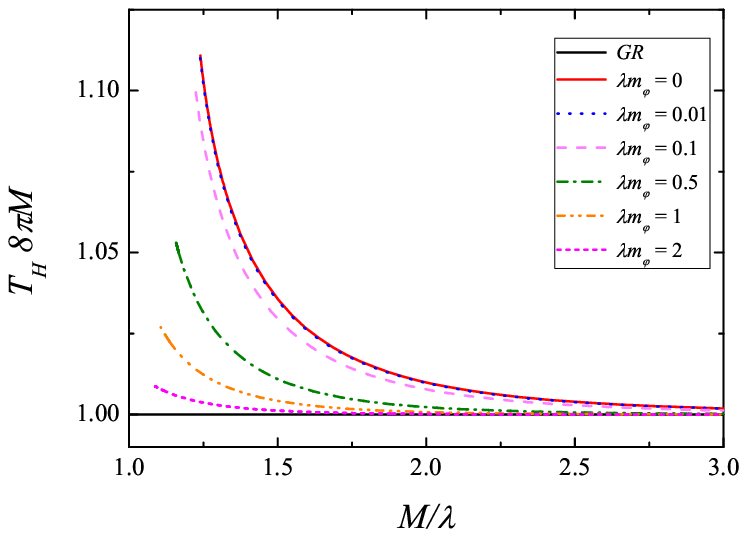}
	\caption{\textit{Left} The temperature of the black hole versus its mass for the linear coupling function \eqref{eq:linear_coupling}. Both are rescaled to the coupling constant. \textit{Right} The normalized to the Schwarzschild limit temperature on the horizon, $T_H 8\pi M$, versus the rescaled mass of the black hole. The notations on both panels are the same as in the above figures. }
	\label{Fig:Th_phi}
\end{figure}

In Fig. \ref{Fig:Th_phi} in the \textit{left} panel we plot the temperature of the black hole versus its mass. The temperature is higher than the GR case for models with low mass, and it gets closer to the GR one with the increase of the mass of the black hole. As one can expect, the temperature converge to the GR one with the increase of the mass of the scalar field. In the \textit{right} panel for better presentation of the deviations from GR we plot the normalized to the Schwarzschild limit temperature $T_H 8\pi M$.

\subsection{Black holes with exponential coupling}

Here we present our numerical results for EdGB black holes with a massive scalar field and coupling function given  by 
\begin{equation}\label{eq:exponential_coupling}
f(\varphi)=  \frac{1}{4} e^{2\gamma\varphi}
\end{equation} 
with $\gamma$ being a parameter. 

\begin{figure}[]
	\centering
	\includegraphics[width=0.45\textwidth]{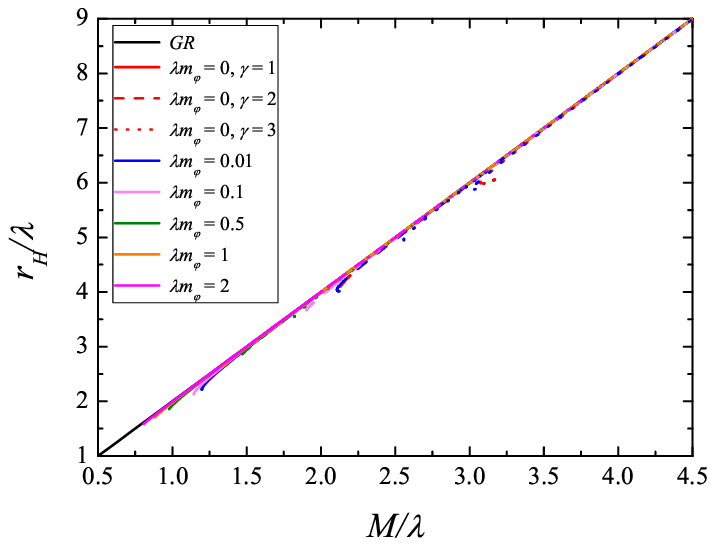}
	\includegraphics[width=0.45\textwidth]{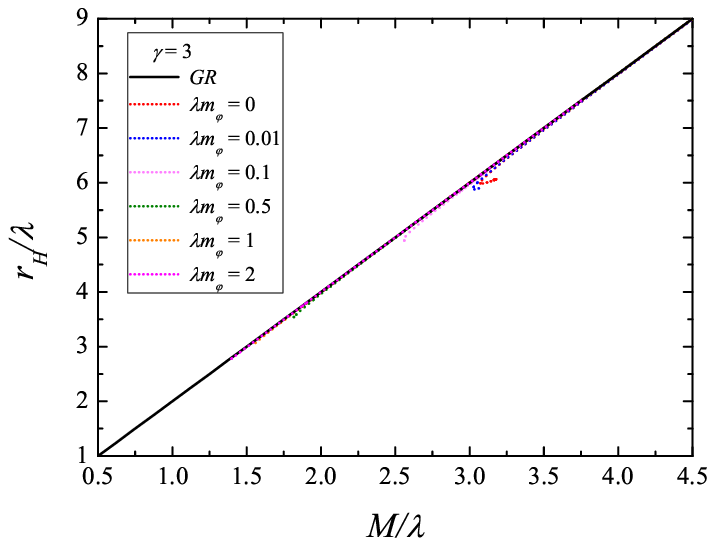}
	\caption{\textit{Left} Radius of the horizon versus the mass of the black hole  for different values of the parameter $\gamma$ for the exponential coupling function \eqref{eq:exponential_coupling}. Both are rescaled with respect to the coupling constant $\lambda$. \textit{Right} Radius of the horizon versus the mass of the black hole for $\gamma = 3$. The notations on both panels are identical. The results are for different values for the rescaled, with the coupling constant, mass of the scalar field $\lambda m_{\varphi}$ in different colours and patterns. }
	\label{Fig:rh}
\end{figure}

In Fig. \ref{Fig:rh} we present the radius of the black hole versus its mass for different masses of the scalar field. In the \textit{left} we present the results for three different values for the parameter $\gamma$, namely $\gamma = 1, \gamma = 2$, and $\gamma = 3$, and in the \textit{right} panel only for $\gamma = 3$. As one can see the behavior is qualitatively the same for all values of $\gamma$ and increasing $\gamma$ leads to an increase of the threshold mass below which no EdGB black holes exist. In the massless limit we could observe the existence of a secondary branch of solutions after a minimum of the mass is reached for all values of $\gamma$, even though in this resolution only the secondary branches for $\gamma=3$ are clearly visible. In the \textit{right panel} only the results for $\gamma=3$ are presented in order to have better visibility of the $m_\varphi$ dependence of the results. One can see that with the increase of the mass of the field the secondary branch gradually disappears (it gets shorter). At the same time the minimal mass for which the EdBG solutions exist significantly shifts to lower masses and it tends to $0$ in the $m_\varphi \rightarrow \infty$ limit. 

\begin{figure}[]
	\centering
	\includegraphics[width=0.45\textwidth]{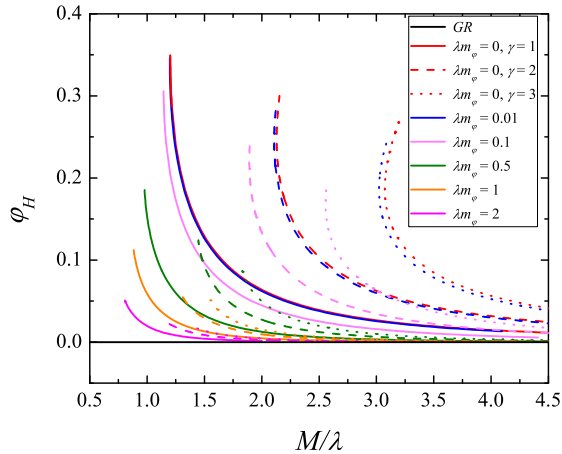}
	\caption{The value of the scalar field on the horizon versus the rescaled mass of the black hole  for the exponential coupling function \eqref{eq:exponential_coupling}.}
	\label{Fig:phi_h}
\end{figure}

In Fig. \ref{Fig:phi_h} we plot the value of the scalar field on the horizon as a function of the black hole mass. In this case as well the value of the scalar field decreases with the increase of the black hole mass, as well as with the increase of the scalar field mass. In the case of $\gamma = 3$ one can clearly see that a minimum of the mass is reached that marks the existence of a second branch of solutions.  

\begin{figure}[]
	\centering
	\includegraphics[width=0.45\textwidth]{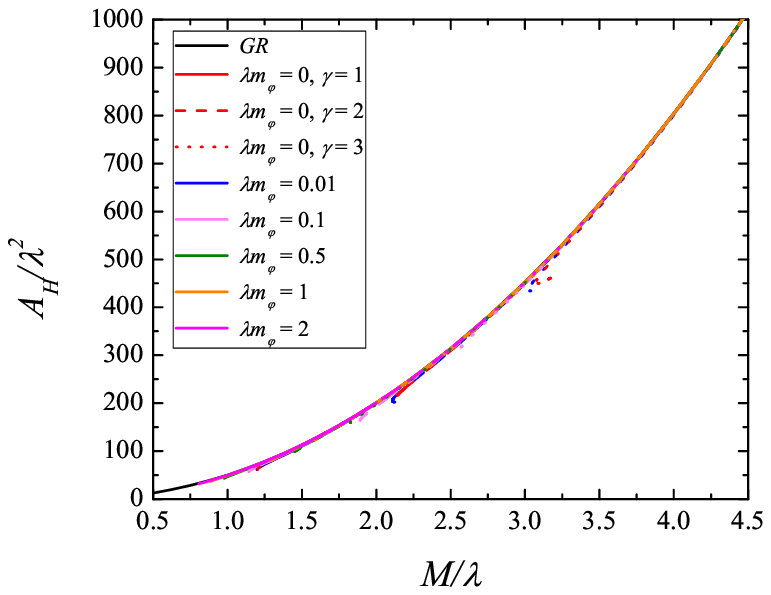}
	\includegraphics[width=0.45\textwidth]{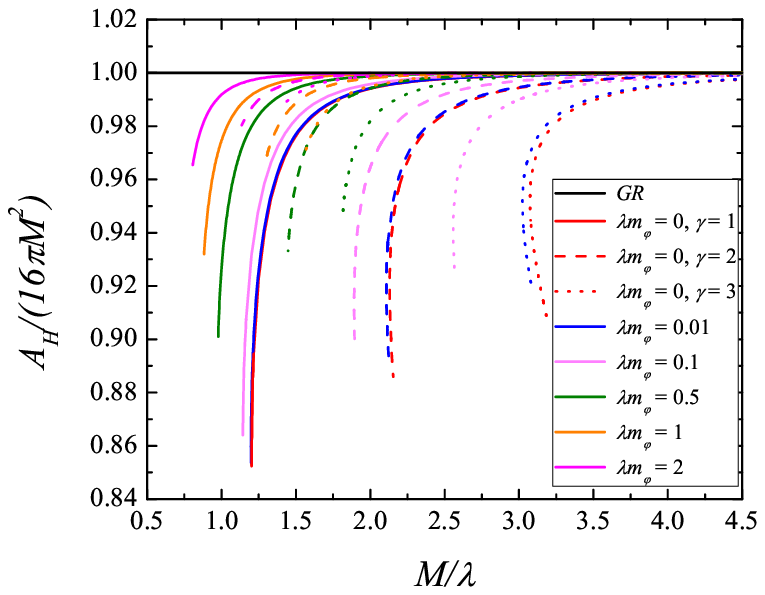}
	\caption{ \textit{Left} The area $A_H$ of the black hole horizon versus the mass of the black hole  for the exponential coupling function \eqref{eq:exponential_coupling}. Both are rescaled with respect to the coupling constant $\lambda$. \textit{Right} The normalized, to the Schwarzschild limit, arena of the horizon $A_H/(16\pi M^2)$ versus the rescaled mass of the black hole. The notations on both panels are identical. The results are for different values of the parameter $\gamma$ and different values for the rescaled, with the coupling constant, mass of the scalar field $\lambda m_{\varphi}$  in different colours and patterns. }
	\label{Fig:Ah}
\end{figure}

In Fig. \ref{Fig:Ah} we plot the area of the horizon versus the black hole mass in the \textit{left} panel, and the normalized area $A_{H}/(16\pi M^2)$ versus the mass in the \textit{right} panel. In the \textit{left} panel one can see that the area of the black hole horizon is smaller than the Schwarzschild one, and it converges to the GR one with the increase of the scalar field mass. The secondary branches are clearly distinguishable in this case as well and the behavior is the one we have  already mentioned -- the second branch of solutions gradually disappears with the increase of the scalar field mass.  In the \textit{right} we print the normalized area of the horizon versus the mass of the black hole. This graph can be used for better representation of how the deviations from GR change with the mass of the scalar field and with the parameter $\gamma$.

\begin{figure}[]
	\centering
	\includegraphics[width=0.45\textwidth]{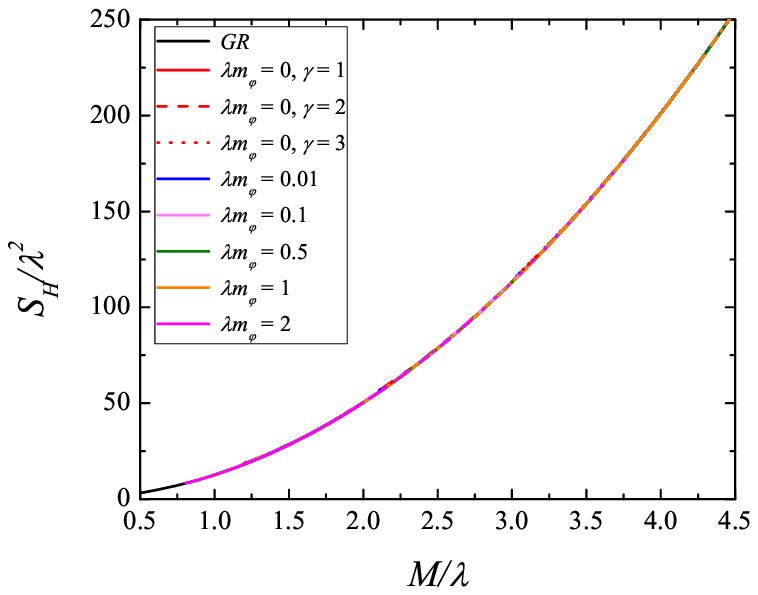}
	\includegraphics[width=0.45\textwidth]{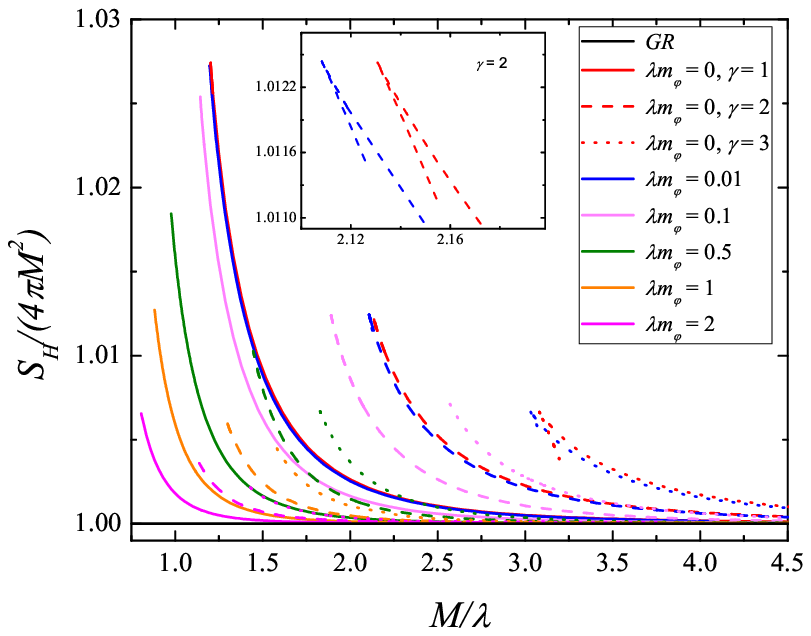}
	\caption{\textit{Left} The entropy of the black hole versus its mass  for the exponential coupling function \eqref{eq:exponential_coupling}. Both are rescaled to the coupling constant. \textit{Right} The normalized to the Schwarzschild limit entropy $S_H/(4\pi M^2)$ on the horizon versus the rescaled black hole mass. The notations are the same as in the figures above.}
	\label{Fig:Sh}
\end{figure}

In the \textit{left} panel of Fig. \ref{Fig:Sh} we plot the entropy on the black hole horizon, calculated via the expression 
\begin{equation}
S_H = \frac{1}{4}A_H + 4 \pi\lambda^2 \left(f(\varphi_{H}) - \frac{1}{4}\right).
\end{equation}
No significant deviations from GR are observed for the  range of masses for which solutions exist and for the scale used in the graphs. In the \textit{right} panel we plot the normalized entropy, $S_H/(4\pi M^2)$, versus the black hole mass in order to study in more details the deviations of the entropy from the GR case. The differences with GR are smalled, compared to the linear coupling. However, the secondary branch of solutions is clearly visible on the figure for high values of $\gamma$ and low values of the scalar field mass (part of the $\gamma = 2$ case is zoomed in the small panel). For all of the studied cases (both massless and massive) the secondary branch has lower entropy compared to the main branch which leads to the conclusion that it is most probably unstable similar to the massless case \cite{Blazquez-Salcedo:2017txk} (see also \cite{Blazquez-Salcedo:2016enn}).

\begin{figure}[]
	\centering
	\includegraphics[width=0.45\textwidth]{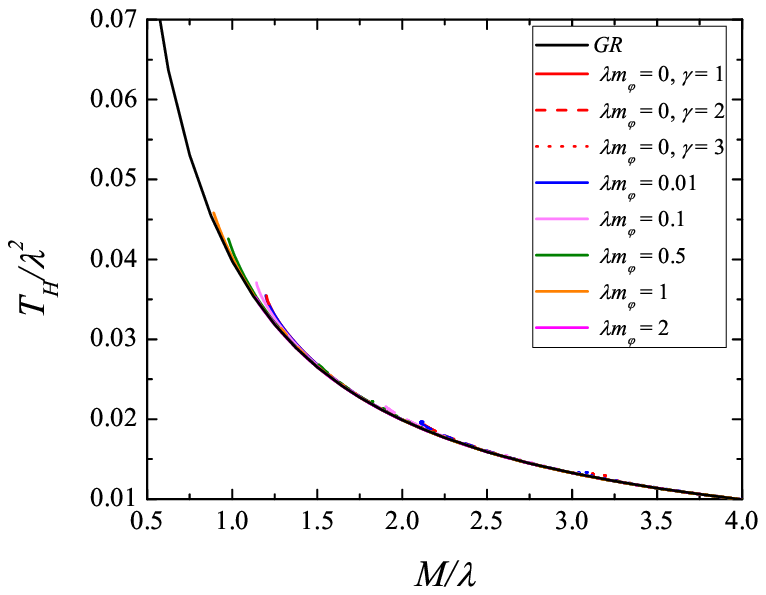}
	\includegraphics[width=0.45\textwidth]{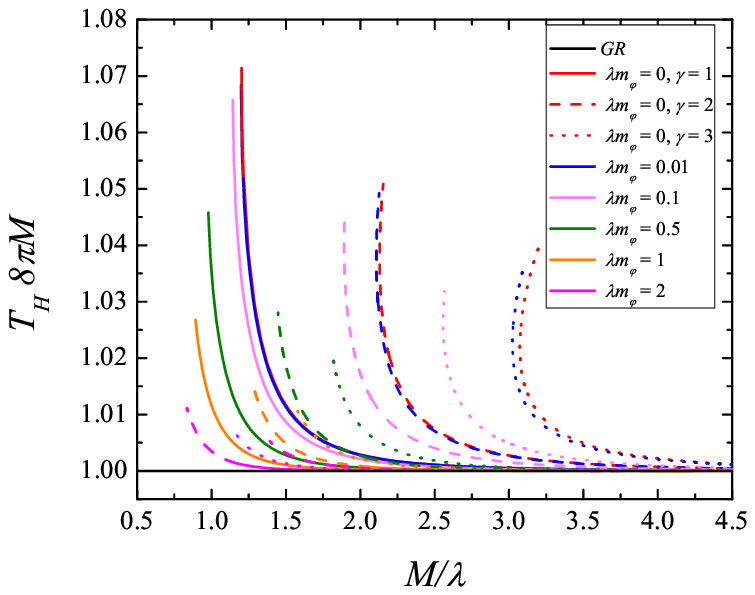}
	\caption{\textit{Left} The temperature of the black hole versus its mass  for the exponential coupling function \eqref{eq:exponential_coupling}. Both are rescaled to the coupling constant. \textit{Right} The normalized to the Schwarzschild limit temperature $T_H 8\pi M$ on the horizon versus the rescaled black hole mass. The notations are the same as in the figures above.}
	\label{Fig:Th}
\end{figure}

In Fig. \ref{Fig:Th} in the \textit{left} panel we plot the temperature of the black hole versus its mass. The temperature is higher than the GR case for models with low mass, and it get closer to the GR one with the increase of the mass of the black hole. As one can expect, the temperature converges to the GR one with the increase of the mass of the scalar field. The secondary branches exist for all three values of $\gamma$ just like in the figures above, but they are not visible due to the scale of the figure.  In the \textit{right} panel for better presentation of the deviations from GR we plot the normalized to the Schwarzschild limit temperature $T_H 8\pi M$. In this case the secondary branches can be distinguished in the figure for high values for $\gamma$.

\subsection{Spontaneously scalarized black holes}

Here we present our numerical results for ESTGB black holes with a massive scalar field and coupling function given  by 

\begin{equation}\label{eq:CoupingScalarization}
f(\varphi)=  \frac{1}{2\beta} \left(1- e^{-\beta\varphi^2}\right)
\end{equation} 
with $\beta>0$ being a parameter. This coupling function in the case when the scalar field is massless allows for a spontaneous scalarization of the Schwarzschild black hole \cite{Doneva_2018}. In our numerical solutions we shall use $\beta=12$ but similar results are observed for other values of $\beta$ as well. We refer the reader to \cite{Doneva_2018a} for an extensive discussion of the influence of the parameter $\beta$ on the properties of the scalarized solutions in the massless case.

As we know from the massless scalar field case \cite{Doneva_2018,Doneva_2018a}, for such coupling in addition to the Schwarzschild solution additional scalarized solutions  branch off  the GR one that can be labeled with the number of zeros of the scalar field. As the stability analysis of the solutions shows, though,  only the solution with no zeros of the scalar field  is stable, while the rest possessing one or more zeros in radial direction are always unstable \cite{Salcedo_2018}. For this reason, in the present paper we will present only the solutions with no zeros of the scalar field. We should note, that the Schwarzschild solutions is also unstable for black hole masses below the mass corresponding to the bifurcation point \cite{Doneva_2018a}. 

\begin{figure}[]
	\centering
	\includegraphics[width=0.45\textwidth]{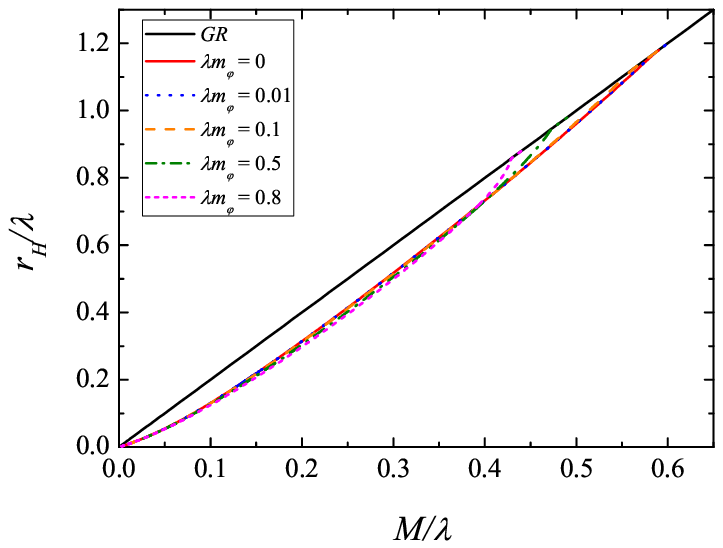}
	\includegraphics[width=0.45\textwidth]{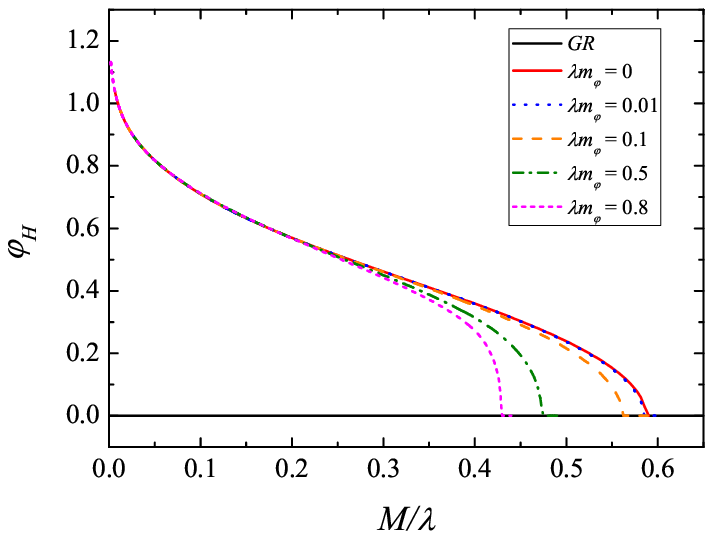}
	\caption{\textit{Left} The radius of the horizon versus the black hole mass  for the scalarized solutions. Both are rescaled with respect to the coupling constant $\lambda$. \textit{Right} The value of the scalar field on the horizon versus the rescaled black hole mass. The notations on both panels are identical. The results are for different values for the scaled mass of the scalar field $\lambda m_{\varphi}$ in different colours and patterns. }
	\label{Fig:rh_scalarized}
\end{figure}

\begin{figure}[]
	\centering
	\includegraphics[width=0.45\textwidth]{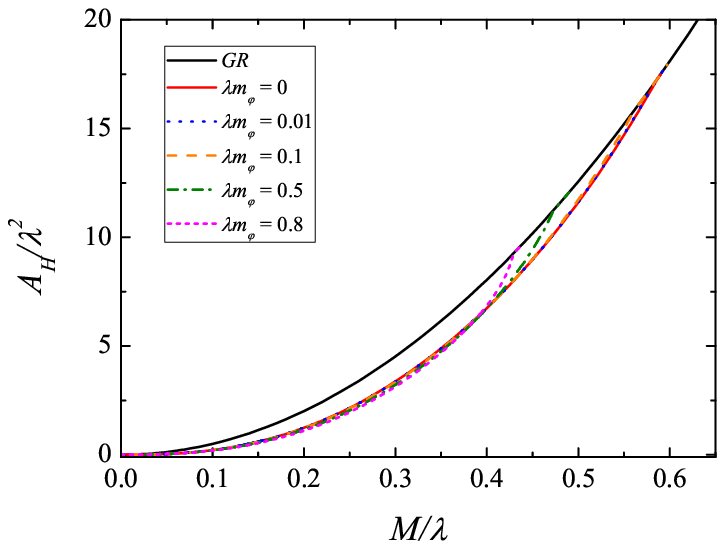}
	\includegraphics[width=0.45\textwidth]{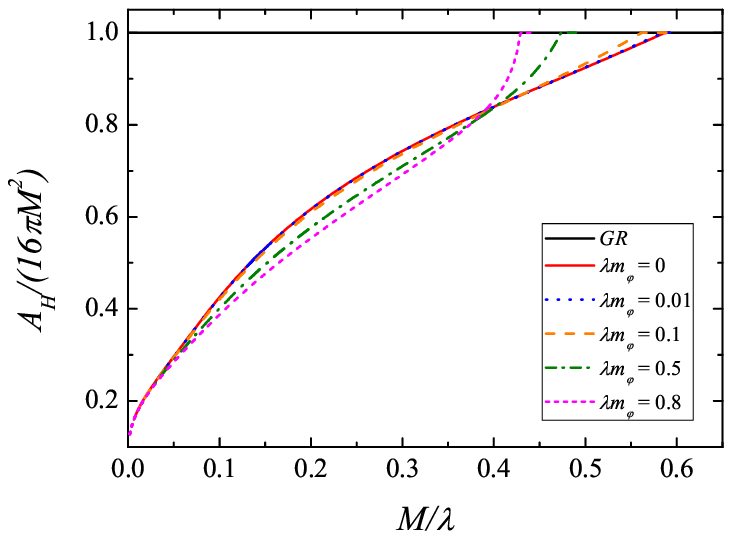}
	\caption{\textit{Left} The area $A_H$ of the black hole horizon versus the mass of the black hole   for the scalarized solutions. Both are rescaled with respect to the coupling constant $\lambda$. \textit{Right} The normalized, to the Schwarzschild limit, arena of the horizon $A_{H}/(16\pi M^2)$ versus the rescaled mass of the black hole. The notations in both panels are identical. The results are for different values for the scaled mass of the scalar field $\lambda m_{\varphi}$  in different colours and patterns. }
	\label{Fig:Ah_scalarized}
\end{figure}

The points of bifurcation for different masses of the scalar field can be best observed in Fig. \ref{Fig:rh_scalarized} where the horizon radius and the scalar field on the horizon are plotted as functions of the black hole mass. In the figure we have plotted only the branches with $\varphi_H>0$. Because of the $Z_2$ symmetry of the dimensionally reduced field equations for our coupling function \eqref{eq:CoupingScalarization} and the potential of the scalar field \eqref{eq:Potential}, it is clear that solutions with an opposite sign of the scalar field but with the same metric functions exist. As one can see the scalar field mass causes significant deviations from the case with a  massless scalar field  mainly for larger black hole masses close to the bifurcation point. For small  black hole masses the solutions with zero and nonzero $m_\varphi$ are almost indistinguishable from the  $m_\varphi=0$ case. The point of bifurcation moves to smaller black hole masses with the increase of $m_\varphi$. The shift of the bifurcation point when a nonzero $m_\varphi$ is introduced was actually first reported and examined in detail for scalarized neutron stars in \cite{Staykov:2018hhc}. This shift of the bifurcation point and the deviations from the massless scalar field case for larger black hole masses can also be observed in Fig. \ref{Fig:Ah_scalarized} where the area of the horizon as a function of the black hole mass is plotted. 

\begin{figure}[]
	\centering
	\includegraphics[width=0.45\textwidth]{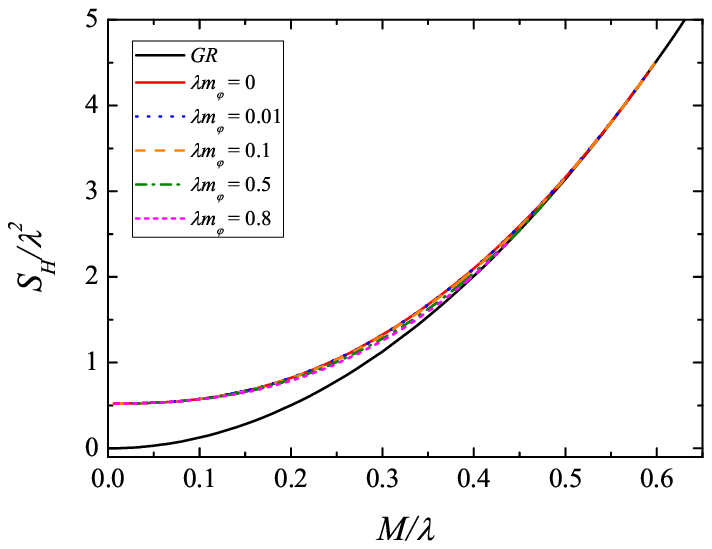}
	\includegraphics[width=0.45\textwidth]{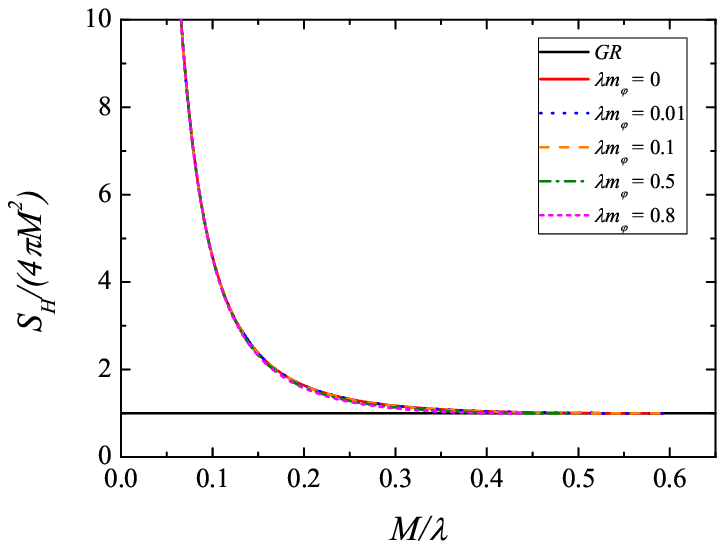}
	\caption{\textit{Left} The entropy of the black hole versus its mass   for the scalarized solutions. Both are rescaled to the coupling constant. \textit{Right} The normalized to the Schwarzschild limit entropy on the horizon, $S_H/(4\pi M^2)$, versus the rescaled mass of the black hole. The notations on both panels are the same as in the figures above. }
	\label{Fig:Sh_scalarized}
\end{figure}

The entropy is calculated using the same formula as for the linear coupling \eqref{eq:S_linear_coupling} and it is plotted in Fig. \ref{Fig:Sh_scalarized} for two different normalizations. As one can see, the scalarized  black hole solutions always have larger entropy compared to the Schwarzschild solution and thus they are thermodynamically favorable. This observation is true both for the massless and the massive branches of solutions. 

The temperature of the scalarized black holes is always larger than the Schwarzschild one. For large black hole masses close to the bifurcation point the temperature in the massive scalar field case is smaller than the massless one and this behavior changes with the increase of the black hole mass.  

\begin{figure}[]
	\centering
	\includegraphics[width=0.45\textwidth]{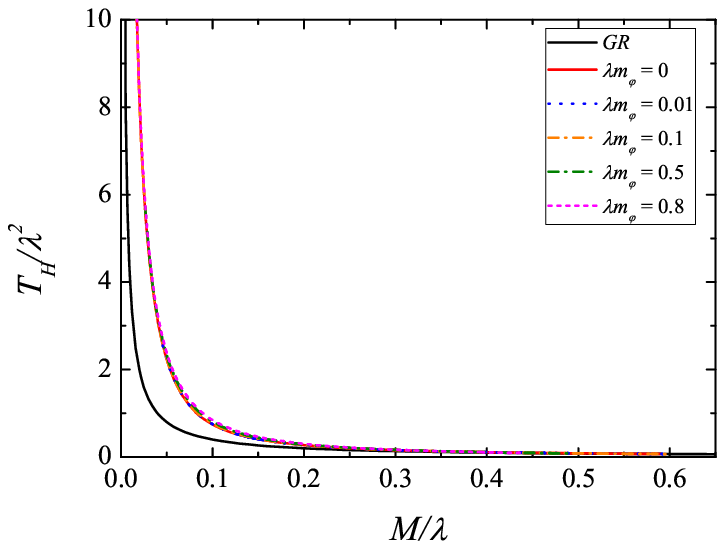}
	\includegraphics[width=0.45\textwidth]{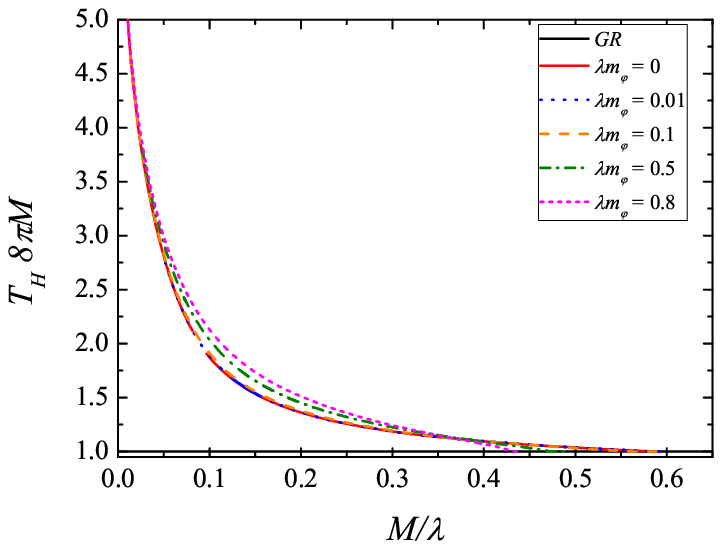}
	\caption{\textit{Left} The temperature of the black hole versus its mass   for the scalarized solutions. Both are rescaled to the coupling constant. \textit{Right} The normalized to the Schwarzschil limit temperature on the horizon, $T_H 8\pi M$, versus the rescaled mass of the black hole. The notations on both panels are the same as in the above figures. }
	\label{Fig:Th_scalarized}
\end{figure}

Finishing this section let us briefly comment on the following. It was recently shown in \cite{Anson_2019} that within the ESTGB theories exhibiting scalariztion  a similar effect may occur in a cosmological background, resulting in the instability of cosmological solutions. In particular, a catastrophic instability could develop during inflation within a period of time much shorter than the minimum required duration of inflation\footnote{In \cite{Anson_2019} it is assumed  that the scalarization field $\varphi$ is not the inflaton
	field. The case when the scalar field $\varphi$ itself is the inflanton field is studied in \cite{Chakraborty:2018scm}.}. As a result, the standard cosmological dynamics is not recovered.  A possible resolution of this problem was very recently proposed in \cite{Macedo:2019sem}. According to the authors of \cite{Macedo:2019sem} the adding mass and quartic self-interaction term for the scalar field could suppress the  tachyonic instability during the inflation. In our opinion the problem with the cosmological instability requires deeper
investigation. In the present paper we consider the ESTGB model exhibiting scalarzation   as an effective model operating only on astrophysical scales without pretending for being a complete theory explaining the early Universe and the dark energy problem. We should also remember that the inflation is just a hypothesis  not firmly established scientific fact and in general 
the inflation itself can not be criteria for adopting or rejecting certain models.

\section{Conclusion}
In the present paper we have studied an extension of the black holes in massless scalar-tensor-Gauss-Bonnet gravity, namely the inclusion of a scalar field mass. Thus, a additional length scale of the problem was introduced and roughly speaking the scalar field is confined within  the Compton wavelength.   Such an effect can reconcile the theory with the observations for a larger range of parameters compared to the massless scalar field case similar to the standard massive scalar-tensor theories. 

We focused on three different standard forms of the coupling function -- a linear coupling between the scalar field and the Gauss-Bonnet invariant, an exponential one and a coupling function which leads to spontaneous scalarization. In the first two cases we have similar qualitative behavior both for massless and for massive scalar field with the important difference that for exponential coupling a secondary branch of black hole solutions for small masses is observed.

We have examined different properties of the black hole solutions -- the horizon radius and area, the entropy and the black hole temperature. For the linear and the exponential coupling the differences with the Schwarzschild solutions are negligible for large black hole masses and they increase with the decrease of the black hole mass. Introducing a scalar field mass has the effect of suppressing the scalar field which brings the sequences of Gauss-Bonnet black holes closer to the Schwarzschild solution which are recovered in the limit when the scalar field mass tends to infinity. For a fixed black hole mass the radius of the horizon decreases with respect to the Schwarzschild solution. 

In the case of linear and exponential coupling function, as we know from the massless scalar field case, there is a threshold black holes mass below which black holes in EdGB gravity do not exist. The inclusion of a scalar field mass leads to a decrease of this threshold mass and thus the domain of existence of solutions is increased.  The second branch of solutions observed in the massless case for exponential coupling  shrinks with the increase of the scalar field mass and for large enough masses of the scalar field we could no longer observe it. We have studied the entropy of the solutions as well. For all of the studied cases the EdGB black holes have larger entropy compared to the Schwarzschild black holes. More importantly, the entropy of the secondary branch of EdGB solutions is smaller than the entropy of the primary branch and this does not change with the introduction of a scalar field mass, which points towards the conclusion that the secondary branch is most probably unstable.
 
In the case of spontaneous scalarization the scalar field mass alters significantly the point of scalarization bringing it to lower masses with the increase of the scalar field mass  and therefore,  the domain of existence of scalarized black hole shrinks. The black holes both in the massive and the massless case have larger entropy compared to the Schwarzschild solution for the considered coupling function. Therefore, they are thermodynamically the preferred solutions over the GR ones. 

 A possible extension of the studies in the present paper is to explore models with various self-interaction terms similar to \cite{Staykov:2018hhc}. An interesting venue of investigation is to consider neutron stars in Gauss-Bonnet gravity with massive and self-interacting scalar field. Such studies are underway.
 
\vskip 0.5cm

\textit{Note added.} During the final editing of the present paper, a preprint studying the spontaneous scalarization of black holes in Gauss-Bonnet gravity with a massive and self-interacting term for a different coupling function appeared on the arXiv \cite{Macedo:2019sem}.

\section*{Acknowledgements}
DD would like to thank the European Social Fund, the Ministry of Science, Research and the Arts Baden-Wurttemberg for the support. DD is indebted to the Baden-Wurttemberg Stiftung for the financial support of this research project by the Eliteprogramme for Postdocs.
KS and SY acknowledge financial support by the Bulgarian NSF Grant KP-06-H28/7. Networking support by the COST Action  CA16104 is also gratefully acknowledged.
%%%%%%%%%%%%%%%%%%%%%%%%%%%%%%%%%%%%%%%%%%%%%%%%%%%%%%%%%%%%%%%%%%%%%%%%%%%%%%%

\end{document}